\documentclass[pra,aps,amsmath, amssymb, twocolumn, superscriptaddress]{revtex4-1}

\usepackage{graphicx}
\usepackage{hyperref}
\usepackage{physics}
\usepackage{color}
\usepackage{enumerate}
\usepackage{bm}
\usepackage[normalem]{ulem}
\usepackage{comment}

\usepackage[caption=false]{subfig}

\definecolor{burntorange}{rgb}{0.8, 0.33, 0.0}

\usepackage{graphicx}
\usepackage{braket}
\usepackage{amsmath, upgreek}

\usepackage{amsthm}
\usepackage{amsfonts, amssymb}
\usepackage{bbm}
\usepackage{hyperref}

\usepackage{mathtools}
\usepackage{mathrsfs}

\DeclarePairedDelimiterX{\infdivx}[2]{(}{)}{%
	#1\;\delimsize\|\;#2%
}

\newcommand*{\Scale}[2][4]{\scalebox{#1}{$#2$}}%

\begin{document}

\newtheorem{theorem}{Theorem}[section]
\newtheorem{proposition}{Proposition}[section]
\newtheorem{corollary}{Corollary}[section]

\theoremstyle{definition}
\newtheorem{definition}{Definition}[section]
\newtheorem{assumptions}{Assumptions}[section]
\newtheorem{example}{Example}[section]
\newtheorem{rmk}{Remark}[section]
\newtheorem{conj}{Conjecture}[section]

\newcommand{\Det}{\text{Det}}
\newcommand{\ergo}{\mathcal{E}}
\newcommand{\deltaoff}{\Delta^{\rm off}}
\newcommand{\en}{E}
\newcommand{\pass}{^{\downarrow}}
\newcommand{\rh}{\hat\rho}
\newcommand{\ham}{\hat{H}}
\newcommand{\hilb}{\mathcal{H}}

\newcommand{\sigmax}{{\hat{\sigma}^x}}
\newcommand{\sigmay}{{\hat{\sigma}^y}}
\newcommand{\sigmaz}{{\hat{\sigma}^z}}

\title{Optimal local work extraction from bipartite quantum systems in the presence of Hamiltonian couplings}

\author{Raffaele Salvia}
\email[]{raffaele.salvia@sns.it}
\affiliation{Scuola Normale Superiore, I-56127 Pisa, Italy}

\author{Giacomo De Palma}
\affiliation{Department of Mathematics, University of Bologna, 40126 Bologna, Italy}

\author{Vittorio Giovannetti}
\affiliation{NEST, Scuola Normale Superiore and Istituto Nanoscienze-CNR, I-56127 Pisa, Italy}

\begin{abstract}
We investigate the problem of finding the local analogue of the ergotropy, that is the maximum work that can be extracted from a system if we can only apply local unitary transformation acting on a given subsystem.
 In particular, we provide a closed formula for the local ergotropy in the special case in which the local system has only two levels, and give analytic lower bounds and semidefinite programming upper bounds for the general case.
As non-trivial examples of application, we compute the local ergotropy for a atom in an electromagnetic cavity with Jaynes-Cummings coupling, and the local ergotropy for a spin site in an XXZ Heisenberg chain, showing that the amount of work that can be extracted with an unitary operation on the coupled system can be greater than the work obtainable by quenching off the coupling with the environment before the unitary transformation.
\end{abstract}

	\maketitle
	
\section{Introduction} \label{sec:intro}
As quantum technologies are expected to be highly sensitive to the interaction with the environment, it is often useful to explicitly include the environment in the modeling of a quantum process. 
One way of representing open quantum systems is to extend the Hilbert space of the quantum system of interest, regarding it as a subspace of a larger Hilbert space which includes the ``environment'' of the system. 
	
The problem of work extraction from quantum systems embedded in an environment was first studied in Ref.~\cite{Frey2014}, which introduced the concept of strong local passivity, or \emph{CP-passivity}. A quantum state is CP-passive with respect to a given subsystem if its energy can not be decreased with ant completely positive map on the subsystem. As the energy extractable with local CPTP map can be found with a ``semidefinite program'' optimization, Ref. \cite{Alhambra2019} provided an algorithm for computing it, and charachterized the necessary and sufficient conditons for CP-local passivity.

As CPTP maps constitute the most general evolution than a quantum system can undergo, the extractable energy studied in~\cite{Frey2014} and~\cite{Alhambra2019} represents an ultimate upper bound on the energy that can be drained from a system using only local operations on a given subsystem. However, just like in the global case~\cite{Niedenzu2019}, it makes sense to consider the energy extractable under a more limited set of allowed operations.  

In \cite{Mukherjee2016} it has been found the maximum energy extractable from a composite system, if all the subsystems are coupled with heath baths at inverse temperature $\beta$. This is the local analogue of the non-equilibrium free energy~\cite{Esposito2011}.
In this work we shall consider instead the local analogue of the ergotropy~\cite{Allahverdyan2004}, that is the energy that we can extract from an isolated system using only local unitary transformations.
In analogy with the notion of (globally) passive states~\cite{Pusz1978, Lenard1978}, one can define the concept of \emph{locally passive} states~\cite{Sen2021}, i.e. states whose energy can not be decreased by arbitrary local unitary transformations. In Ref.~\cite{Sen2021} it was found a set of necessary and sufficient conditions for the local passivity, but only in the case in which the Hamiltonian of the system is a sum of local Hamiltonians.
We can define the \emph{local ergotropy} with respect to a subsystem $S$ as the energy extractable using local unitary operations\cite{Satoya2022} on $S$.
Since the problem of finding local ergotropy can not be expressed as a SDP (as the unitarity constraint $UU^\dagger = I$ is not linear), it has been considerably less studied. In general, any correlated system exhibits an \emph{ergotropic gap}~\cite{Mukherjee2016, Alimuddin2019}, meaning that the local ergotropy is strictly smaller than the global ergotropy, or that correlations with the environment are detrimental for work extraction \cite{OPPE2002, VITA2019, GOOLD2017, BERA2017, MANZ2019, ANDO2019}. 
On the contrary,  the role of initial correlations among the various subsystem can be beneficial when extracting work via global unitary operations~\cite{HUBER2015,FRANCICA2017,NOSTRO}).

It is important to stress that in all  the works mentioned above the global Hamiltonian of the  systems is always assumed to be interaction-free. Exceptions to this general trend can be found in Refs.~\cite{BARRA,STARSB,DECHIARA,ANDOLINA2018,ITO} where, studying energy exchanges  not directly related with ergotropy calculations, 
the presence of couplings among the various subsystems is taken into consideration by adding to the energy bill  the cost associated with the  abrupt switching-off and switching-on of  such terms. 
Apart from these works, it seems however that no general study of the local ergotropy has been presented when the model explicitly exhibits 
coupling among the various  subsystems.  
The aim of this paper is to fill this gap.
In the case in which the local system of interest is a two-level system we present a simple method to exactly compute the maximal amount of work one can extract locally from a correlated many-body quantum system for Hamiltonian models which explicitly exhibit coupling terms among the various sub-systems. 
In the case in which the system has a bigger dimension, we present some general bounds for its local ergotropy.
 
The material is organized as follows:  In Sec.~\ref{sectionI}  we formalize the notion of local ergoropy  and draw some connection with previous literature; 
 In Sec.~\ref{sec:metodo_generale} we describe a general optimization method to compute the local ergotropy which takes a considerably simpler form in the case in which the system $S$ is a qubit;
In Sec.~\ref{sec:applicazioni} we apply our technique to compute the local ergotropy of two simple systems: an atom in an optical cavity with Jaynes-Cummings coupling and of a site in an  anisotropic (XXZ) Heisenberg spin chain.  In both systems, we find regimes in which the local ergotropy is bigger than the work that can be extracted by decoupling the system from its environment (i.e., of the ergotropy of the decoupled local system, minus the energetic cost of isolating the system). Conclusions and outlooks are finally presented in Sec.~\ref{conclusione}. 

\section{Local ergotropy} \label{sectionI} 

Consider a bipartite quantum system $SE$ intialiazed in the possibly correlated quantum state 
$\hat{\rho}_{SE}$ and characterized by the joint Hamiltonian 
\begin{eqnarray}
\label{scomposizione_hamiltoniana}
\ham_{SE} :=  \ham_S \otimes \hat{I}_E + \hat{I}_S \otimes \ham_E + \hat{V}_{SE} \;,
\end{eqnarray}
with $\ham_S$ and $\ham_E$ being the local energy terms and with $\hat{V}_{SE}$ being the interaction contribution
which we assume to have zero partial trace on the $S$ side, i.e.  $\mbox{Tr}_S[\hat{V}_{SE}]=0$.
The (global) \emph{ergotropy} \cite{Allahverdyan2004} of the state $\rh_{SE}$ with respect to the Hamiltonian $\ham_{SE}$ is the maximum amount of energy that can be extracted from $\rh_{SE}$ by means on unitary transformation acting on the bipartite system $SE$; in formula it is expressed by the positive-semidefinite functional 
\begin{eqnarray}
\label{def_ergotropy}
&&\ergo (\rh_{SE}, \ham_{SE})   \\
&:=&\max_{\hat{U}_{SE} \in \mathbf{U}(d_Sd_E)} \left\{ \Tr[\ham_{SE}(
\rh_{SE} -   \hat{U}_{SE} \rh_{SE} \hat{U}^\dagger_{SE}) ]\right\}  \nonumber \\
&=&\Tr[\rh_{SE}\ham_{SE}] - \min_{\hat{U}_{SE} \in \mathbf{U}(d_Sd_E)} 
\mbox{Tr}[\hat{U}_{SE} \rh_{SE} \hat{U}^\dagger_{SE} \ham_{SE}]\; . \nonumber 
\end{eqnarray}
To gain insight in the problem~(\ref{def_ergotropy}), it is useful to consider the following classical analogue. Let $S_{\rm cl}$ be a classical system which may be in one of the $d$ states $\{ s_1 , \dots s_d \}$, having energies $\{ \epsilon_1 , \dots \epsilon_d \}$. A \emph{state} of this classical system is specified by a probability distribution $\vec{p} = \{ p_1, \dots p_d \}$ over the $d$ states. The expected value of the energy of the system in the state $\vec{p}$ is $\langle E \rangle = \sum_{i=1}^d p_i \epsilon_i$. We can act on the classical system by applying an arbitrary permutation $\pi \in S_d$ on the states, so that $\vec{p} \to \pi(\vec{p}) = \{p_{\pi(1)}, \dots p_{\pi(d)}  \}$. Then classical analogue of the ergotropy problem is to find the permutation which maximizes the expected value of the energy decrement, that is
\begin{eqnarray}
\ergo_{\rm cl}(\vec{p}, \vec{\epsilon}) = \sum_{i=1}^d p_i \epsilon_i - \min_{\pi \in S_d}  \sum_{i=1}^d p_{\pi(i)} \epsilon_i
\label{classical_ergo}
\end{eqnarray}
The classical problem~\ref{classical_ergo} can be immediately solved using the \emph{rearrangement inequality} \cite{hardy1952inequalities}, which states that
\begin{eqnarray}
\min_{\pi \in S_d}  \sum_{i=1}^d p_{\pi(i)} \epsilon_i =  \sum_{i=1}^d p^\downarrow_{i} \epsilon^\uparrow_i \; ,
\end{eqnarray}
where $p^\downarrow_{i}$ denote the components of the vector $\vec{p}$ arranged in decreasing order (that is, $p^\downarrow_1 \geq \dots \geq p^\downarrow_d$), and similarly $\epsilon^\uparrow_i$ are the energies arranged in increasing order. 
The quantum ergotropy problem~(\ref{def_ergotropy}) can be solved in a completely analogous way, invoking the Hermitian-matrices analogue of the rearrangement inequality, i.e. Von Neumann's trace inequality \cite{Mirsky1975}:
\begin{equation}
\min_{\hat{U}_{SE} \in \mathbf{U}(d_Sd_E)} \mbox{Tr}[\hat{U}_{SE} \rh_{SE} \hat{U}^\dagger_{SE} \ham_{SE}] = \sum_{i=1}^d p^\downarrow_i \epsilon^\uparrow_i \; ,
\end{equation} 
where this time $p^\downarrow_i$ are the eigenvalues of the density matrix $\rh_{SE}$ arranged in decreasing order, and $\epsilon^\uparrow_i$ are the eigenvalues (energy levels) of the Hamiltonian $\ham_{SE}$ arranged in increasing order. Writing $\rh_{SE} = \sum_{i=1}^d p^\downarrow_i \ket{i}_{SE}\bra{i}$ and 
$\ham_{SE} = \sum_{i=1}^d \epsilon^\uparrow_i \ket{\epsilon_i}_{SE}\bra{\epsilon_i}$, the optimal unitary transformation which achieves the minimum is given by
$\hat{U}_{SE}^{({\rm opt})} := \sum_{i=1}^d \ket{\epsilon_i}_{SE}\bra{i}$.

The \emph{$S$-local ergotropy} of the model is now defined as the maximum amount of work one can extract from $SE$ by means of local unitary operations that act locally on $S$ while not affecting $E$, i.e. 
\begin{eqnarray}
\label{def_localergotropy}
&&\ergo_S (\rh_{SE}, \ham_{SE})  \\
&&:=
\max_{\hat{U}_S \in \mathbf{U}(d_S)} \mbox{Tr}[\ham_{SE} ( \rh_{SE} - (\hat{U}_S \otimes \hat{I}_{E}) \rh_{SE} (\hat{U}_S^\dagger \otimes \hat{I}_{E}) ) ]\;,\nonumber
\end{eqnarray}
where the maximization is performed over the set $\mathbf{U}(d_S)$ of the unitary transformations on
the $d_S$-dimensional Hilbert space associated to $S$. 
This is a non-negative quantity 
which by construction is upper-bounded by $\ergo(\rh_{SE}, \ham_{SE})$.
Simple algebra reveals that $\ergo_S (\rh_{SE}, \ham_{SE})$ bares no functional dependence upon the local Hamiltonian of $H_E$ of the subsystem $E$ and that it is convex with respect to $\rh_{SE}$  and $\ham_{SE}$.  We observe  that 
in the absence of interactions 
 (i.e. for $\hat{V}_{SE}=0$),  the $S$-local ergotropy reduces to the ergotropy $\ergo(\rh_S, \ham_S):= \max_{\hat{U}_S \in \mathbf{U}(d_S)}  \mbox{Tr}[\ham_{S} ( \rh_{S} - \hat{U}_S \rh_{S} \hat{U}^\dagger_S ) ]$ of the reduced state  $\rh_S:=\mbox{Tr}_E[\rh_{SE}]$   associated with the local Hamiltonian $\ham_S$, i.e. 
 \begin{eqnarray} \label{freecase} 
\ergo_S (\rh_{SE}, \ham_{SE})\Big|_{\hat{V}_{SE}=0}    &=& \ergo(\rh_S, \ham_S)\;,
\end{eqnarray}
 which in the case where $\hat{V}_{SE}\neq 0$ is  sufficiently regular,  can be replaced by the inequality 
  \begin{eqnarray} \label{freecaseineq} 
\Big| \ergo_S (\rh_{SE}, \ham_{SE})-   \ergo(\rh_S, \ham_S)\Big|  \leq 2 \| \hat{\rho}_{SE}\|_2  \| \hat{V}_{SE}\|_2\;,
\end{eqnarray}
with $\| \hat{\Theta} \|_2= \sqrt{\mbox{Tr}[ \hat{\Theta}^\dag \hat{\Theta}]}$ representing the Hilbert-Schmidt norm of the operator $\hat{\Theta}$ (see Appendix~\ref{APPEPRIM}). 
Notice also that when the input state of the system factorizes $\rh_{SE} = \rh_{S}\otimes \rh_{E}$, $\ergo_S (\rh_{SE}, \ham_{SE})$ reduces to the erogotropy
of the density $\rh_{S}$ evaluated for the effective free local Hamiltonian 
$\ham^{(eff)}_S : = \ham_S + \mbox{Tr}_E[  \hat{V}_{SE} \rh_E]$ obtained by adding to $\ham_S$ the interaction term contracted on the state of $E$, 
i.e. 
 \begin{eqnarray}
\ergo_S (\rh_{S}\otimes \rh_{E}, \ham_{SE})  &=& \ergo(\rh_S, \ham^{(eff)}_S)\;.
\end{eqnarray} 
  Beside Eq.~(\ref{freecaseineq})  no universal ordering can be drawn between $\ergo_S (\rh_{SE}, \ham_{SE})$ and 
 $\ergo(\rh_S, \ham_S)$. Similar considerations also apply if we compare $\ergo_S (\rh_{SE}, \ham_{SE})$
 with the work $\ergo_S^{\rm off}(\rh_{SE}, \ham_{SE})$ one can get in a two stage procedure where 
  first the coupling term $\hat{V}_{SE}$ is abruptly switched-off as in Ref.~\cite{BARRA,STARSB,DECHIARA,ANDOLINA2018,ITO}, and then local operations are applied to 
  resulting interaction-free Hamiltonian model. As discussed in Appendix~\ref{appendA}, by neglecting Lamb-shift  corrections this quantity can be estimated as
  \begin{eqnarray}
\label{def_localergotropy_off}
\ergo_S^{\rm off}  (\rh_{SE}, \ham_{SE})  := \ergo(\rh_S, \ham_S) - 
\Delta^{\rm off}(\rh_{SE},\hat{V}_{SE})\;,
\end{eqnarray}
with 
\begin{eqnarray}
\label{def_deltaoff}
\Delta^{\rm off}(\rh_{SE},\hat{V}_{SE}) :=- \mbox{Tr}[\rh_{SE} \hat{V}_{SE}] \; ,
\end{eqnarray}
being the energy cost associated with the switching-off event.
 In this case Eq.~(\ref{freecaseineq}) gets replaced by the inequality
  \begin{equation} \label{freecaseineq12} 
\Big| \ergo_S (\rh_{SE}, \ham_{SE})-  \ergo_S^{\rm off}  (\rh_{SE}, \ham_{SE})\Big|  \leq  \| \hat{\rho}_{SE}\|_2  \| \hat{V}_{SE}\|_2\;,
\end{equation}
which while bounding 
the distance between    $\ergo_S  (\rh_{SE}, \ham_{SE})$
 and $\ergo_S^{\rm off}  (\rh_{SE}, \ham_{SE})$ cannot be used to establish a general ordering among them.

	\section{General formulas and bounds}
	\label{sec:metodo_generale}
The study of the local ergotropy functional~(\ref{def_localergotropy}) is considerably more complex than the global one~(\ref{def_ergotropy}). To appreciate this fact, consider the local analogue of the classical rearrangement problem~(\ref{classical_ergo})
\begin{eqnarray}
\ergo_{S, \rm{cl}}(\vec{p}, \vec{\epsilon}) = \sum_{i,j} p_{i,j}\epsilon_{i,j} - \min_{\pi}  \sum_{i,j} p_{{\pi(i)},j}\epsilon_{i,j} \; ,
\label{classical_localergo}
\end{eqnarray}
obtained by considering the case in which given a probability distribution  $p_{i,j}$ on a set of bipartite classical states $\{ s_{i,j} \}$ specified by two indices $i$ and $j$, 
and characterized by energies $\epsilon_{i,j}$, one is asked to improve the mean energy of the model  
by permuting only one of two indices  (i.e. sending $p_{i,j} \to p_{\pi(i),j}$).
The minimization  in~(\ref{classical_localergo}) is an istance of the ubiquitous and widely studied \emph{assignment problem}~\cite{Monge1784, burkard2009assignment}. It can be efficiently solved with several algorithms~\cite{Jacobi1865, Konig1931, Bertsekas1988}, but the solution can not be written with a closed formula in terms of $\{ p_{i,j} \}$ and $\{ \epsilon_{i,j} \}$. Since the quantum problem~(\ref{def_localergotropy}) includes as a special case the classical problem~(\ref{classical_localergo}) (which can be seen as the case in which $\rh_{SE}$ and $\ham_{SE}$ are both diagonal in a tensor product basis), this implies that no general closed solution can exist for 
$\ergo_S (\rh_{SE}, \ham_{SE})$.
 Even the set of states $\rh$ for which $\ergo_S(\rh, \ham) = 0$ does not admit an easy characterization, and it is not in general a convex set, in contrast which the sets of states such that $\ergo(\rh, \ham) = 0$, which is a simplex in the space of density matrices\cite{PerarnauLlobet2015}.
 As we shall see in the forthcoming subsections, an explicit formula can however be derived
 in the special case where the quantum system $S$ has only two levels. Furthermore 
a bound for $\ergo_S (\rh_{SE}, \ham_{SE})$ can be obtained in terms of 
the maximum energy decrement under local unital transformation, which can be calculated with a semidefinite programming (SDP) optimization (see appendix~\ref{sec:bounds}).

\subsection{A closed formula for a single qubit} 
To get a closed expression for the $S$-local ergotropy
	we find it useful to adopt the 
  Generalized Pauli Operators (GOPs) expansion formalism 
 reviewed in Appendix~\ref{sec:prelim}.
 In the case where $S$ is a finite-dimensional system, this allows us to represent the density matrix 
 $\rh_{SE}$ as 
 	\begin{eqnarray}
	\hat\rho_{SE} &=&\frac{\hat{I}_S}{d_S} \otimes \hat\rho_{E} + \frac{1}{2}\sum_{i=1}^{d_S}  \hat\sigma_{S}^{(i)} \otimes    \hat\rho_{E}^{(i)}\;, 
 \label{decomposizione_statooperatore}
	\end{eqnarray}
with
$\hat\rho_{E}:= \mbox{Tr}_S[ \hat\rho_{SE}]$ the reduced state of $E$, $ \hat\rho_{E}^{(i)}:= \mbox{Tr}_S[\hat\sigma_{S}^{(i)} \hat\rho_{SE}]$,  $\{ \hat\sigma_{S}^{(i)} ; i=1,\cdots, d_S^2-1\}$ 	being the  GPO set adopted for $S$. 
 Similarly we can write 
 the Hamiltonian terms $\ham_{S}$ and $\hat{V}_{SE}$ as 
	\begin{eqnarray}\label{decomposizione_hamiltoniana_loc}
\ham_{S} &=& c \; \hat{I}_S+ \sum_{i=1}^{d^2_S-1} h_i \; \hat\sigma_{S}^{(i)}   \;, \\ 
\hat{V}_{SE} &=& \frac{1}{2} 
\sum_{i=1}^{d^2_S-1} \; \hat\sigma_{S}^{(i)}  \otimes  \hat{V}_{E}^{(i)} \;, 
\label{decomposizione_hamiltoniana}
\end{eqnarray}
with $c: = \mbox{Tr}[\ham_S]/d_S$, ${h}_i := 
\mbox{Tr}[\hat\sigma_{S}^{(i)}  \hat{H}_S]/2$, and $ \hat{V}_{E}^{(i)}:= \mbox{Tr}_S
[\hat\sigma_{S}^{(i)}  \hat{V}_{SE}]$. Notice that in writing Eq.~(\ref{decomposizione_hamiltoniana}) we assumed that $\hat{V}_{SE}$ contains no expansion term that is proportional to the identity: as a matter of fact, in case such
term exists we can drop it by properly redefine $\ham_E$~\cite{NOTA1}. 
Invoking now the orthonormal conditions of the GPO (see Eq.~(\ref{traccia_pauli})), 
 we can express Eq.~(\ref{scomposizione_ergo}) in the compact form
 \begin{eqnarray}
\ergo_S (\rh_{SE}, \ham_{SE})= \max_{\hat{U}_S \in \mathbf{U}(d_S) } \mbox{Tr}[
{\cal O}_U {\cal M} - {\cal M} ] \;, 
\label{Epass_Bloch2}
\end{eqnarray}
with  ${\cal O}_U \in {\bf O}(d_S^2-1)$ being the orthogonal matrix which corresponds to the unitary $\hat{U}_S$ in the selected GOP representation and 
with ${\cal M}$ the $(d_S^2-1)\times (d_S^2-1)$ real matrix of elements 
\begin{eqnarray}
\label{def_minf}
\mathcal{M}_{ik} := - \left( r_i h_{k}+ \frac{1}{2} \mbox{Tr}_E [   \hat{\rho}_{E}^{(i)} \; \hat{V}_{E}^{(k)} ] \right)\;,
\end{eqnarray}
with $r_i:= \mbox{Tr}[ \hat\sigma_{S}^{(i)}  \rh_S]$ the components of the generalized Bloch vector of
the reduced density matrix $\rh_S$. 
The solution which maximizes the right-hand-side of Eq.~(\ref{Epass_Bloch2}) has, in general, no closed formula expression; but it can be solved with a convex optimization algorithm (e.g., steepest descent).
An exception to this is provided by the special case where $S$ is a qubit, i.e. for  $d_S = 2$ \cite{NOTA2}.  
Under these circumstances in fact one has that 
$\{ {\cal O}_U \mid \hat{U}_S \in U(2) \}$ exactly coincides with the
subgroup $SO(3)$ of the orthogonal group $O(3)$. If the matrix $\mathcal{M}$ has an even number of negative eigenvalues - that is, if $\mbox{det}[ {\cal M}]\geq 0$, we can find by polar decomposition a matrix in $SO(3)$ which turns $\mathcal{M}$ into $|{\cal M}|$. If instead $\mathcal{M}$ has an odd number of negative eigenvalues ($\mbox{det}[ {\cal M}]< 0$), no special orthogonal matrix can transform it into $|{\cal M}|$, and the best that one can do is to transform $\mathcal{M}$ into a matrix whose smallest eigenvalue is negative. Therefore we can write 
 \begin{eqnarray}
&&\ergo_S (\rh_{SE}, \ham_{SE})\Big|_{\rm qubit}=\max_{{\cal O} \in SO(3) } \mbox{Tr}[
{\cal O} {\cal M} - {\cal M} ] \nonumber \\
&&\quad =  \left\{ \begin{array}{ll} 
\mbox{Tr}[ |{\cal M}| - {\cal M}] \;, & \mbox{for $\mbox{det}[ {\cal M}]\geq 0$\;,} \\ \\ 
\mbox{Tr}[ |{\cal M}| - {\cal M}] - \frac{2}{\| {\cal M}^{-1}\|}\;,  & \mbox{for $\mbox{det}[ {\cal M}]< 0$\;,} 
\end{array} \right.
\label{Epass_Bloch2_qubit}
\end{eqnarray}
with $|{\cal M}| := \sqrt{{\cal M}^\dag {\cal M}}$ and $\| \cdots\|$ representing the operator norm.

\subsection{Bounds for $d_S\geq 3$}
\label{sec:bounds}
In this section we derive some bounds for the local ergotropy functional   $\ergo_S (\rh_{SE}, \ham_{SE})$ in the case of $d_S>2$. 
\\

\paragraph{Polar upper bound:--} 
To begin with  let us observe  that 
the set of orthogonal matrices $\{ {\cal O}_U \mid \hat{U}_S \in U(d_S) \}$ is a proper subgroup in 
the subgroup $SO(d_S^2-1)$ it follows that the formula in the right-hand-side  of Eq.~(\ref{Epass_Bloch2_qubit}) is always a proper upper bound for
$\ergo_S (\rh_{SE}, \ham_{SE})$, i.e. 
\begin{eqnarray}
&&\ergo_S (\rh_{SE}, \ham_{SE})\leq\max_{{\cal O} \in SO(d_S^2-1) } \mbox{Tr}[
{\cal O} {\cal M} - {\cal M} ] \nonumber \\
&&\quad =  \left\{ \begin{array}{ll} 
\mbox{Tr}[ |{\cal M}| - {\cal M}] \;, & \mbox{for $\mbox{det}[ {\cal M}]\geq 0$\;,} \\ \\ 
\mbox{Tr}[ |{\cal M}| - {\cal M}] - \frac{2}{\| {\cal M}^{-1}\|}\;,  & \mbox{for $\mbox{det}[ {\cal M}]< 0$\;.} 
\end{array} \right.
\label{ineq1}
\end{eqnarray}
Saturation of the inequality~(\ref{ineq1}) is unlikely unless the  matrix ${\cal M}$ admits polar decomposition
\begin{eqnarray}
{\cal M} = {\cal O}_* |{\cal M}| \;, 
\end{eqnarray} 
with the orthogonal matrix ${\cal O}_*$ being an element of  $\{ {\cal O}_U \mid \hat{U}_S \in U(d_S) \}$ (a fact that always occurs
for $d_S=2$).
\\ 

\paragraph{SDP upper bound:--}
An alternative bound for 
the local ergotropy~(\ref{def_localergotropy}) can be obtained by exploiting convexity 
argument to write 
\begin{eqnarray} \label{relaxation_to_convex}
&&\min_{\hat{U}_S \in \mathbf{U}(d_S)} \Tr\left[ \ham_{SE} (\hat{U}_S \otimes \hat{I}_{E}) \rh_{SE} (\hat{U}_S^\dagger \otimes \hat{I}_{E}) ) \right]  \\ \nonumber 
&& \qquad \qquad = \min_{\Phi_S \in \overline{\mathbf{U}(d_S)}} \Tr\left[  \ham_{SE}(\Phi_S \otimes \mathbb{I}_{E}) \left( \rh_{SE} \right)  \right] \; ,
\end{eqnarray}
where $\mathbb{I}_{E}$ represents the identity map on $E$ and where  $\overline{\mathbf{U}(d_S)}$ denotes the set of all convex combinations of unitary channels, i.e., $\overline{\mathbf{U}(d_S)} := \{ \Phi_S \mid \Phi_S(
\hat{\rho}_S) = \sum p_k U_k \hat{\rho}_S U^\dagger_k \; , U_k \in \mathbf{U}(d_S), p_k > 0, \sum p_k = 1  \}$.
Following Ref.~\cite{Alhambra2019} we can now introduce the operator 
\begin{eqnarray} \hat{C}_{SS'} := \Tr_E \left[ \rh^{T_S}_{SE}\;  \ham_{S'E} \right]\;,
\end{eqnarray}  where $\rh^{T_S}_{SE}$ is the partial transpose of $\rh_{SE}$ and recall that for any quantum channel $\Phi_S$ on $S$ it holds that
\begin{eqnarray}
\Tr\left[ \ham_{SE} (\Phi_S \otimes \mathbb{I}_{E}) \left( \rh_{SE} \right)  \right] = \Tr\left[ \hat{C}_{SS'} \hat{E}_{SS'}^{(\Phi_S)} \right] \; ,
\end{eqnarray}
with $\hat{E}_{SS'}^{(\Phi_S)}$ being the Choi matrix~\cite{Choi1975,Jiang2013} of the channel $\Phi_S$.
Accordingly we can express (\ref{def_localergotropy}) 
as
\begin{equation}
\label{locergo2}
\ergo_S (\rh_{SE}, \ham_{SE}) = \Tr[\ham_{SE}\rh_{SE}] - 
\min_{\Phi_S \in \overline{\mathbf{U}(d_S)}} \Tr\left[ \hat{C}_{SS'} \hat{E}_{SS'}^{(\Phi_S)} \right] .
\end{equation}
A computable upper bound can extracted from this by 
relaxing the minimization  to include all the $\Phi_S$ belonging to the set of \emph{unital} channels, $\mathfrak{U}(d_S) \supset \overline{\mathbf{U}(d_S)}$~\cite{Mendl2009}, i.e. the quantum channels such that $\Phi_S({I}_S) ={I}_S$. 
With this relaxation, we obtain a SDP bound for the local ergotropy:
\begin{equation}
\label{relaxation_to_unital}
\ergo_S (\rh_{SE}, \ham_{SE})  \leq \Tr[\ham_{SE}\rh_{SE}] - \min_{\hat{E}_{SS'} } \Tr\left[ \hat{C}_{SS'} \hat{E}_{SS'} \right] \; , \end{equation}
where now the minimum is mow performed over the whole set of operators fulfilling the conditions 
\begin{eqnarray} 
 & \hat{E}_{SS'} \geq 0 \;,  \label{completely_positive} \\
 &  \Tr_S \hat{E}_{SS'} = \hat{I}_{S'} \label{unital} \;, \quad \Tr_{S'} \hat{E}_{SS'} = \hat{I}_{S} \label{trace_preserving} \; , 
\end{eqnarray}
(the first ensuring  complete positivity, the second ensuring trace preservation, and the last the unitality requirement). 
Notice that, in the case $d_S = 2$, we have $\mathfrak{U}(d_S) = \overline{\mathbf{U}(d_S)}$, and therefore the bound~(\ref{relaxation_to_unital}) coincides with the exact formula~(\ref{Epass_Bloch2_qubit}). When $d_S \geq 3$, however, the bound~(\ref{relaxation_to_unital}) will be in general larger than the local ergotropy of the system.

\section{Examples}

\label{sec:applicazioni}

\begin{figure}
	\centering
	\includegraphics[width=\columnwidth]{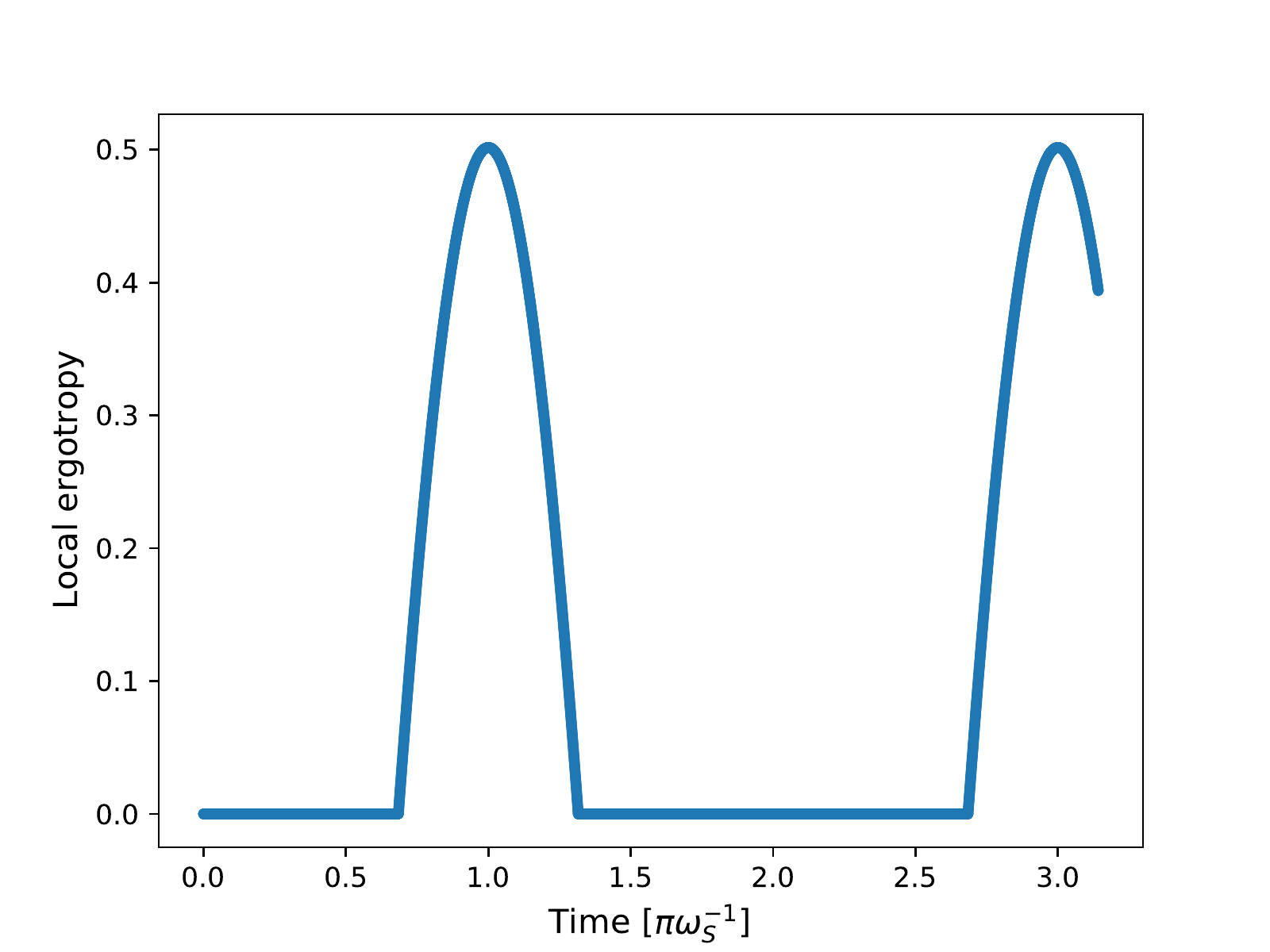}
	\caption{Local ergotropy of the state $\cos\alpha\ket{n,+} + \sin\alpha\ket{n,-} e^{i\Omega t}$ as a function of  the time $t$, for the choice of parameters $\omega_S = 1$, $\omega_E = 1.2$, and $\Omega = 0.1$, $n=10$ and $\alpha=0.4\pi$.}
	\label{fig1_phi}
\end{figure}

The simplest, yet non trivial model of quantum optics is the  Jaynes-Cummings model which describes the interaction of a two-level atom $S$, with energy levels spaced by $ \omega_S$, with the electromagnetic radiation field of a  high-Q cavity mode $E$  of frequency $\omega_E$~\cite{JaynesCummings1963, Shore1993}. Expressed in terms of the two-level atom Pauli operators its Hamiltonian is given by:
\begin{equation}
\label{hamiltoniana_JaynesCummings}
\ham_{SE}^{(\rm JC)}  :=   \omega_E \; \hat{a}^\dagger \hat{a}+ \frac{\omega_S}{2} \hat{\sigma}_{S}^{z} + \frac{\Omega}{2} \left( \hat{a} \otimes \hat{\sigma}_S^+ + \hat{a}^\dagger
\otimes \hat{\sigma}_S^- \right)\;,
\end{equation}
with $\Omega$ the {Rabi frequency}, $a$, $a^\dag$ the annihilation and creation operator of the mode,  
 and  
$\hat{\sigma}_S^+:=( \hat{\sigma}_{S}^{x} - i \hat{\sigma}_{S}^{y})/2$, 
$\hat{\sigma}_S^-:=( \hat{\sigma}_{S}^{x} +  i \hat{\sigma}_{S}^{y})/2$ the  raising and lowering operators of the atom
(hereafter $\hbar=1$). 
The Hamiltonian  $\ham_{SE}^{(\rm JC)}$  admits as energy eigenvectors 
the  (dressed) states 
\begin{eqnarray}
\label{autostati__JaynesCummings}
\ket{n, +} &:=& \cos \theta_n \ket{1} \otimes \ket{n} + \sin \theta_n \ket{0} \otimes \ket{n+1} \;, \\
\ket{n, -} &:=& \sin \theta_n \ket{1} \otimes \ket{n} - \cos \theta_n \ket{0} \otimes \ket{n+1} \;, 
\end{eqnarray} 
with $\ket{n}$ being the $n$-th Fock state of the cavity mode and 
\begin{eqnarray}
\label{def_thetan}
\theta_n := \frac{1}{2} \arctan \left( \frac{\Omega \sqrt{n+1} }{\omega_S - \omega_E} \right) \;,
\end{eqnarray}
the corresponding eigenvalues being
$E_{n,\pm}:= \omega_E  n  \pm \frac{1}{2}  \sqrt{(\omega_S - \omega_E)^2 + \Omega^2(n+1)}$. 
By direct application of Eqs.~(\ref{def_deltaoff}) and (\ref{def_minf}) we get 
\begin{equation}
\label{def_deltaoff1}
\Delta^{\rm off}(\rh_{SE},\hat{V}_{SE}) :=- \frac{ \Omega}{2} \left( \langle\hat{\sigma}_{S}^{x} \otimes \hat{X} \rangle - \langle\hat{\sigma}_{S}^{y} \otimes \hat{Y} \rangle\right) \; ,
\end{equation}
and 
\begin{eqnarray} 
\!\!\!\mathcal{M} = -\frac{1}{2} \left[ \begin{array}{lll}  \Omega\langle\hat{\sigma}_{S}^{x} \otimes \hat{X} \rangle
&-   \Omega\langle\hat{\sigma}_{S}^{x} \otimes \hat{Y} \rangle & \omega_S \langle\hat{\sigma}_{S}^{x} \rangle \\
   \Omega\langle\hat{\sigma}_{S}^{y} \otimes \hat{X} \rangle
& -  \Omega\langle\hat{\sigma}_{S}^{y} \otimes \hat{Y} \rangle &\omega_S \langle\hat{\sigma}_{S}^{y} \rangle\\
   \Omega\langle\hat{\sigma}_{S}^{z} \otimes \hat{X} \rangle
& -   \Omega\langle\hat{\sigma}_{S}^{z} \otimes \hat{Y} \rangle &\omega_S\langle\hat{\sigma}_{S}^{z} \rangle
\end{array} \right],  \label{computeM} 
\end{eqnarray} 
with  $\hat{X}:= (\hat{a}^\dag + \hat{a})/2$, 
$\hat{Y}:= i (\hat{a}^\dag - \hat{a})/2$, and where we used $\langle \cdots \rangle$ to indicate the expectation value w.r.t. $\rh_{SE}$. 
In what follows we shall focus on the special cases where 
the input state $\rh_{SE}$ corresponds to one of the eigenvectors $|n, \pm \rangle$
of the model showing that under such assumption the $S$-local ergotropy of the model are never
smaller than the corresponding switch-off ergotropy values~(\ref{def_localergotropy_off}). 
To see this, let us start observing that
Eq.~(\ref{def_deltaoff1})   gives $\Delta^{\rm off}(|n, \pm \rangle; \hat{V}_{SE}) =0$. Therefore 
using the fact that for $\theta_n$ as in Eq.~(\ref{def_thetan}) one has 
$\cos^2\theta_n\geq \sin^2\theta_n$ we get \begin{equation}
\ergo^{\rm off}_S(\ket{n,+},\ham_{SE}^{(\rm JC)})
= \ergo_S(\rh_S^{(n,+)},\ham_{S}) =
 \omega_S \cos 2\theta_n \; ,   \label{switch} 
 \end{equation}
 which by construction is always positive semidefinite, and 
 \begin{eqnarray} 
 \ergo^{\rm off}_S(\ket{n,-},\ham_{SE}^{(\rm JC)}) &=& \ergo_S(\rh_S^{(n,-)},\ham_{S}) = 0\;,   
\end{eqnarray}
(in the above expressions $\rh_S^{(n,\pm)}$ stand for the reduced density matrices on $S$ of $|n,\pm\rangle_{SE}$).  
Notice next that   replacing $|n, + \rangle$ in Eq.~(\ref{computeM}) we get instead
\begin{eqnarray} 
\Scale[0.9]{
\!\!\!\mathcal{M}_{+} =  \tfrac{1}{2}    \left[ \begin{array}{ccc}  - \Omega \frac{\sqrt{n+1}}{2} \sin 2 \theta_n&0 &
0 \\0 &  \Omega \frac{\sqrt{n+1}}{2} \sin 2 \theta_n & 0 \\
0
& 0 &-\omega_S \cos 2\theta_n 
\end{array} \right] \;,\nonumber}
\end{eqnarray} 
which has
determinant that is always positive semidefinite due to the fact that 
 Eq.~(\ref{def_thetan})  forces $|\theta_n|\leq \pi/4$.
 Therefore in this case  Eq.~(\ref{Epass_Bloch2_qubit}) implies 
 \begin{equation} 
\ergo_S (|n,+\rangle, \ham^{(\rm JC)}_{SE}) = \omega_S \cos 2\theta_n + 
 \frac{\sqrt{n+1}}{2}  | \Omega \sin 2\theta_n |\;, 
\end{equation} 
that is clearly  greater than or equal to the corresponding switch-off value~(\ref{switch}) -- the gap being an increasing function of the intensity of the coupling term and of the index level $n$. 
Similarly assuming  as input state  $|n, -\rangle$, we obtain a matrix $\mathcal{M}_{-} = - \mathcal{M}_{+}$,
whose determinant is now always negative semidefinite. Therefore from Eq.~(\ref{Epass_Bloch2_qubit}) we get \begin{eqnarray} 
 &&\ergo_S (\ket{n,-}, \ham^{(\rm JC)}_{SE})= \tfrac{\sqrt{n+1}}{2}    |\Omega \sin 2\theta_n|\\
&& \quad  \qquad  - \min\{ 
 \omega_S \cos 2\theta_n, 
 \tfrac{\sqrt{n+1}}{2}    |\Omega \sin 2\theta_n|\} 
 \;, \nonumber 
\end{eqnarray} 
which again is always greater than  the corresponding (zero) switch-off value 
$\ergo^{\rm off}_S(\ket{n,-},\ham_{SE}^{(\rm JC)})$ reported in (\ref{switch}) --  the only exception being
the  weak-coupling regime ($\omega_S \cos 2\theta_n\geq 
\tfrac{\sqrt{n+1}}{2}  |\Omega \sin 2\theta_n|$)  where also $\ergo_S (|n,-\rangle, \ham^{(\rm JC)}_{SE})$ nullifies. 
Most notably at resonance ($\omega_S=\omega_E\Rightarrow |\theta_n|=\pi/4$)
 the gap between the local and the switch-off ergotropy terms of $|n,-\rangle_{SE}$  match those recorded for $|n,+\rangle_{SE}$ as one has 
$\ergo^{\rm off}_S(\ket{n,\pm},\ham_{SE}^{(\rm JC)})\Big|_{\rm res} = 0$ and 
$\ergo_S(\ket{n,\pm},\ham_{SE}^{(\rm JC)})\Big|_{\rm res} =  \tfrac{\sqrt{n+1}}{2}    |\Omega|$.
 On the contrary, one notices that  in the off-resonant regime (i.e. for $|\omega_S-\omega_E|\gg |\Omega|\sqrt{n+1}  \Rightarrow |\theta_n|\simeq 0$), for both $|n,+\rangle_{SE}$ and $|n,-\rangle_{SE}$ the  gap between
  $\ergo_S$ and $\ergo^{\rm off}_S$
  always tends to collapse to zero.
It is also worth to notice that, at variance with the global ergotropy, the local ergotropy functional does in general change with the time evolution of the system. In figure~\ref{fig1_phi} we plot, as an example, the local ergotropy of a coherent mixture of the states $\ket{n=10,+}$ and $\ket{n=10,-}$. The system alternates between time intervals in which $\ergo_S = 0$ and intervals in which $\ergo_S > 0$, with a behaviour reminiscent of the entanglement sudden death and revival~\cite{Yu2009, Yonac2006}. 

As a second example assume the system $S$  to be one element  of a XXZ Heisenberg model of  $N$ spin-1/2 particles disposed on a ring~\cite{baxter2007}. In this case  Hamiltonian can be expressed as 
\begin{eqnarray}
\label{hamiltoniana_heisenberg}
&&\ham^{(\rm{XXZ})}_{SE} := \epsilon \sum_{i=1}^{N} \sigmaz_i 
\\ \nonumber 
&&\qquad - \sum_{i=1}^{N} \left[ J(\sigmax_i\otimes \sigmax_{i\oplus1} + \sigmay_i\otimes \sigmay_{i\oplus 1}) + J_z \sigmaz_i\otimes \sigmaz_{i\oplus1} \right]\;, 
\end{eqnarray}
with the positive constants  $\epsilon$, $J$ and $J_z$  representing the local energy contribution and the coupling therms of the model, and where, to enforce periodic boundary conditions, 
$\oplus$ indicates the sum modulus $N$.
We remind that as $\ham^{(\rm{XXZ})}_{SE}$ admits  the total magnetization $\hat{S}^z := \sum_{i=1}^N \sigmaz_i$ as a conserved quantity  ($[\ham^{(\rm{XXZ})}_{SE}, S_z] = 0$), we can diagonalize it on subspaces of fixed values of $\hat{S}_z$. Specifically assuming $J$ to be sub-leading term with respect to $\epsilon$ and $J_z$, the ground state of the model is provided by $E_G:=  -N(\epsilon + J_z)$ corresponding to the all spin down state  $\ket{\phi_G} := \bigotimes_{i=1}^N \ket{\downarrow}_i$ (total magnetization sector with  $\hat{S}^z =-N$). 
The next excited states $\ket{\phi_k}$ can instead be found on the $\hat{S}^z =-N+2$ sector spanned by superpositions of
 vectors 
$\sigmax_n \ket{\phi_0} := \left(\bigotimes_{i=1}^{n-1} \ket{\downarrow}_i \right) \otimes \ket{\uparrow}_n \otimes \left(\bigotimes_{i=n+1}^{N} \ket{\downarrow}_i \right)$ which have $n-1$ spin down and one spin up. 
Specifically invoking the  Bethe ansatz \cite{Bethe1931, Yang1966}  the corresponding eigenvectors of 
$\ham^{(\rm{XXZ})}_{SE}$  can be expressed as 
\begin{equation}
\label{singlyexcitedstates_heisenberg}
\ket{\phi_k} := \frac{1}{\sqrt{N}} \sum_{n=1}^N e^{\frac{2\pi k i}{N} n} \sigmax_n \ket{\phi_0} \; ,
\end{equation}
with $k$ an integer term belonging to the interval $(-\lfloor N/2 \rfloor, \lfloor N/2 \rfloor]$, 
 the associated
 eigenvalues being 
$E_k :=-\left[ (N-2)\epsilon + (N-4)J_z + 4J\cos\left( \frac{2k \pi}{N}  \right) \right]$. 
In what follow we shall compute the local ergotropy and switch-off local ergotropy for these special states. 
To so we notice that identifying $S$ with the spin $1$ of the model and with $E$ the remaining ones, 
from~(\ref{def_deltaoff}) and (\ref{def_minf}) we get \begin{eqnarray}
\nonumber 
\Delta^{\rm off}(\rh_{SE},\hat{V}_{SE}) &=& J\langle\hat{\sigma}^x_{S} \otimes \hat{Y}_E\rangle + J\langle\hat{\sigma}^y_{S} \otimes \hat{X}_E \rangle   \\
&& + J_z \langle\hat{\sigma}^z_{S} \otimes \hat{Z}_E\rangle\;,
\label{def_deltaoff_Heis} \end{eqnarray}
with $\hat{X}_E:= \hat{\sigma}^x_{N} + \hat{\sigma}^x_{2}$,
$\hat{Y}_E:= \hat{\sigma}^y_{N} + \hat{\sigma}^y_{2}$, and $\hat{Z}_E:= \hat{\sigma}^z_{N} + \hat{\sigma}^z_{2}$, and
\begin{equation} 
\Scale[0.8]{
\mathcal{M} = -\frac{1}{2} \left[ \begin{array}{llc}  J\langle\hat{\sigma}^x_{S} \otimes   \hat{X}_E\rangle &J\langle\hat{\sigma}^x_{S} \otimes   \hat{Y}_E \rangle &
 \epsilon \langle\hat{\sigma}^x_S\rangle +  J_z\langle\hat{\sigma}^x_{S} \otimes   \hat{Z}_E \rangle \\
 J\langle\hat{\sigma}^y_{S} \otimes   \hat{X}_E \rangle
&J\langle\hat{\sigma}^y_{S} \otimes  \hat{Y}_E \rangle &
 \epsilon \langle\hat{\sigma}^y_S\rangle + J_z\langle\hat{\sigma}^y_{S} \otimes  \hat{Z}_E \rangle \\
 J\langle\hat{\sigma}^z_{S} \otimes  \hat{X}_E  \rangle
& J\langle\hat{\sigma}^z_{S} \otimes   \hat{Y}_E\rangle &
 \epsilon \langle\hat{\sigma}^z_S\rangle + J_z\langle\hat{\sigma}^z_{S} \otimes   \hat{Z}_E\rangle\\
\end{array} \right]
} \;.  \label{computeM_Heis} 
\end{equation} 
Taking hence as $\rh_{SE}$ the pure state (\ref{singlyexcitedstates_heisenberg}) 
this yields 
\begin{equation}
\label{deltaoff_Heis}
\Delta^{\rm off}(\ket{\phi_k}_{SE},\hat{V}_{SE}) = \frac{8J}{N} \cos\left( \tfrac{2\pi k}{N} \right) + \frac{2N - 8}{N} J_z \;,
\end{equation} 
and
\begin{equation}
\mathcal{M}_k = \begin{bmatrix}
\tfrac{2J}{N}\cos\left( \frac{2\pi k}{N} \right) & 0 & 0 \\
0 & \tfrac{2J}{N}\cos\left( \tfrac{2\pi k}{N} \right) & 0 \\
0 & 0 &  \frac{N-1}{N}\epsilon +  \tfrac{N-4}{N} J_z \end{bmatrix} \;.
\nonumber \\
\label{M_Heisenberg}
\end{equation}
Notice next that the reduced density matrix $\rh_S$ of the first spin is given by 
$\hat{\rho}_S = \tfrac{N-1}{N} \ket{0}_S\bra{0} + \tfrac{1}{N}\ket{1}_S\bra{1}$
which is passive with respect to the local Hamiltonian $\ham_S= \epsilon \hat{\sigma}_S^z$ of the model.
Accordingly from Eq.~(\ref{def_localergotropy_off}) we get 
  \begin{equation}
\label{def_localergotropy_off-XXZ}
\ergo_S^{\rm off}  (\ket{\phi_k}_{SE}, \ham^{(\rm{XXZ})}_{SE})  = - \frac{8J}{N} \cos\left( \tfrac{2\pi k}{N} \right) - \frac{2N - 8}{N} J_z,
\end{equation}
which can be positive for small $N$ (i.e. $N\leq 4$) and $|k|\geq N/4$, while being always negative in the large $N$ limit. 
Regarding the local ergotropy we treat here explicitly the case where $\frac{N-1}{N}\epsilon +  \tfrac{2N-8}{N} J_z\geq 0$
 which from Eq.~(\ref{Epass_Bloch2_qubit}) 
allows us to write 
\begin{equation}
	\ergo_S(\ket{\phi_k};\ham^{(\rm{XXZ})}_{SE}) =
	\begin{cases}
		0 & \left( \lvert k \rvert \leq N/4 \right) \\
		\frac{8J}{N}| \cos\left( \tfrac{2\pi k}{N} \right)|   & \left( \lvert k \rvert > N/4 \right) \;.
	\end{cases}  
\label{LocalErgotropy_Heis}
\end{equation}
We can hence recognize that as long as $N\geq 4$, $\ergo_S(\ket{\phi_k};\ham^{(\rm{XXZ})}_{SE})$ is always greater or equal to the corresponding 
switch-off value. Exactly the opposite occurs instead for small ($N<4$) rings  as long as the ratio between $J$ and $J_z$ is small enough to ensure the
applicability of~(\ref{LocalErgotropy_Heis}):  under these conditions in fact, for $\lvert k \rvert \leq N/4$ we have $\ergo_S(\ket{\phi_k},\ham^{(\rm{XXZ})}_{SE})=0$,
while $\ergo_S^{\rm off}  (\ket{\phi_k}_{SE}, \ham^{(\rm{XXZ})}_{SE})$ gets positive.

\section{Conclusions} \label{conclusione} 
We derived an exact closed formula for the \emph{local ergotropy} of a two-level system, or the maximum work that can be extracted with local unitary operations from said system interacting with a general environment.
We have shown two examples in which the local ergotropy is strictly bigger than the amount of work that can be obtained by first isolating the system, and then performing the unitary operation.  This indicates that the environment may be a resource, and not only a nuisance, for work extraction.

The formula also gives a (loose) upper bound for the local ergotropy of a system of generic dimension $d$.
The problem of finding the local ergotropy of a system of dimension $d$ can be seen as a quantum generalization of the assignment problem, hence we know that no general closed formula can exist for its solution. However, a more careful analysis may improve the bounds provided here.


\section*{Acknowledgement}
We thank P. Faist for suggesting the relaxation of the problem to local unital channels, which provides an SDP-computable upper bound to the local ergotropy,  and G. M. Andolina and M. Perarnau-Llobet for discussions and comments. We acknowledge financial support  by MIUR (Ministero dell’Istruzione, dell’Universit{\'a} e della Ricerca) by PRIN 2017 “Taming complexity via Quantum Strategies: a Hybrid Integrated Photonic approach” (QUSHIP) Id. 2017SRN- BRK, and via project PRO3 Quantum Pathfinder. GDP is a member of the ``Gruppo Nazionale per la Fisica Matematica (GNFM)'' of the ``Istituto Nazionale di Alta Matematica ``Francesco Severi'' (INdAM)''

\bibliographystyle{ieeetr}
\bibliography{ErgoLocale}

\begin{thebibliography}{10}
\bibitem{NOTA1}
If $S$ and $E$ are finite dimensional systems this assumption can be made
  without loss of generality. Indeed given $\hat{V}_{SE}$ whose partial trace
  with respect to $S$ is not null we can express $\ham_{SE}$ in the desired
  form by simply redefining the coupling term as $\hat{V}'_{SE} = \hat{V}_{SE}
  - \hat{I}_S\otimes \mbox{Tr}_S[ \hat{V}_{SE}]/d_S$ and the local energy
  contribution on $E$ as $\ham_{E}'= \ham_{E}+ \mbox{Tr}_S[ \hat{V}_{SE}]/d_S$
  (here $d_S$ is the dimension of the Hilbert space of $S$).
\bibitem{NOTA2}
An explicit formula can also be derived when the Hamiltonian $\ham_S +
  \hat{V}_{SE}$ has only two level and $\rh_{SE}$ is pure (see
  Appendix~\ref{fine}).
\end{thebibliography}

\newpage
\appendix
\section{Derivation of Eq.~(\ref{freecaseineq})} \label{APPEPRIM} 
Indicating with $\hat{U}^{(\rm{free})}_S$ the unitary transformation that 
enters in the computation of the local ergotropy in the non interacting case~(\ref{freecase}), we can write
\begin{eqnarray}  \nonumber 
&&\ergo_S (\rh_{SE}, \ham_{SE}) \geq\\
&&\quad  \mbox{Tr}[\ham_{SE} ( \rh_{SE} - (\hat{U}^{(\rm{free})}_S \otimes \hat{I}_{E}) \rh_{SE} (\hat{U}_S^{(\rm{free})^\dagger} \otimes \hat{I}_{E}) ) ]\nonumber  \\ 
&& \quad =  \ergo_S  (\rh_{S}, \ham_{S})  + \mbox{Tr}\left[(\rh_{SE} -\hat{U}^{(\rm{free})} \rh_{SE}
\hat{U}^{(\rm{free})\dag}) \hat{V}_{SE}\right] \nonumber  \\
&&  \quad \geq   \ergo_S  (\rh_{S}, \ham_{S})  - \| (\rh_{SE} -\hat{U}^{(\rm{free})} \rh_{SE}
\hat{U}^{(\rm{free})\dag}) \|_2 \| \hat{V}_{SE}\|_2 \nonumber \\ 
&&  \quad \geq   \ergo_S  (\rh_{S}, \ham_{S})  -2 \| \rh_{SE}\|_2 \| \hat{V}_{SE}\|_2\;, \label{DER1} 
\end{eqnarray}
where in the third passage we applied the Cauchy-Schwarz inequality of the Hilbert-Smith scalar product. 
On the contrary by decomposing the maximization of~(\ref{def_localergotropy}) into two independent  maximizations that involve the free and interaction terms of  $\ham_{SE}$ respectively, we can write 
maximization 
\begin{eqnarray} 
\!\!\!\ergo_S (\rh_{SE}, \ham_{SE}) &\leq& \ergo_S  (\rh_{S}, \ham_{S})  + 
\ergo_S (\rh_{SE}, \hat{V}_{SE}) \nonumber \\ 
&\leq& \ergo_S  (\rh_{S}, \ham_{S})  + 2 \| \rh_{SE}\|_2 \| \hat{V}_{SE}\|_2\;. \label{DER2} 
\end{eqnarray}

\section{Local energy extraction for abrupt switching-off of the coupling terms} 
\label{appendA} 
Here we present the derivation of Eq.~(\ref{def_localergotropy_off}) by estimating the maximum work we can extract from $\rh_{SE}$ in the two-steps scenario where, through a quench we first abruptly switch-off the coupling between $S$ and $E$ and then use local unitaries $\hat{U}_S$. 
At the beginning of the process the mean energy contained in the model is given by the expectation value
\begin{eqnarray}
E_{0} &:=& \mbox{Tr}[ \rh_{SE} \ham_{SE}]\nonumber \\
&=& \mbox{Tr}[ \rh_{S}\ham_{S} ] + \mbox{Tr}[ \rh_{E}\ham_{E} ]
+ \mbox{Tr}[ \rh_{SE} \hat{V}_{SE}]\;.
\end{eqnarray}  The switch-off procedure transforms the initial Hamiltonian $\ham_{SE}$ into an interaction-free term of the form
\begin{eqnarray}
\label{scomposizione_hamiltoniana_free}
\ham_{SE}^{\rm free}  =  \ham'_S \otimes \hat{I}_E + \hat{I}_S \otimes \ham'_E  \;.
\end{eqnarray}
Notice that in principle, due to the presence of Lamb-shift contributions, the new local terms $\ham'_S$, $\ham'_E$ 
need not to coincide with the corresponding values $\ham_S$, $\ham_E$  appearing in Eq.~(\ref{scomposizione_hamiltoniana}). Determining the exact structure of $\ham'_S$, $\ham'_E$ strongly 
depends on the specific physical model we are considering. Typically however one expects 
the discrepancies between $\ham'_S$, $\ham'_E$  and $\ham_S$, $\ham_E$ 
to be small, and for the sake of simplicity in our analysis we neglect them.  Accordingly we evaluate the energy
of the system immediately after the quench as 
\begin{eqnarray} 
E_{1} &:=& \mbox{Tr}[ \rh_{SE}\ham_{SE}^{\rm free} ] =
\mbox{Tr}[ \rh_{S}\ham_{S} ] + \mbox{Tr}[ \rh_{E}\ham_{E} ] \nonumber \\
&=& E_0 + \ergo_S^{\rm off}  (\rh_{SE}, \ham_{SE}) \;,
\end{eqnarray} 
where in the second line we invoked Eq.~(\ref{def_localergotropy_off}). 
When positive, the difference between $E_1$ and $E_0$ (i.e. the quantity
$\ergo_S^{\rm off}  (\rh_{SE}, \ham_{SE})$)  accounts for the energy we need to provide to the system
$SE$ in order to suppress the interactions between the subsystems. Such term has hence to be subtracted from the maximal work 
we can extract from $\rh_{SE}$ via local unitary on $S$ in the second part of the protocol, i.e. the quantity 
\begin{eqnarray}
\nonumber 
&&\ergo_S (\rh_{SE}, \ham^{\rm free}_{SE})  \\
&&:=
\max_{\hat{U}_S \in \mathbf{U}(d_S)} \mbox{Tr}[\ham_{SE}^{\rm free} \left( \rh_{SE} - (\hat{U}_S \otimes \hat{I}_{E}) \rh_{SE} (\hat{U}_S^\dagger \otimes \hat{I}_{E}) \right) ] \nonumber \\
&&= \ergo (\rh_{S}, \ham_{S})  \;. \label{def_localergotropy_free}
\end{eqnarray}
Equation~(\ref{def_localergotropy_off}) then simply follows by putting together these observations.

The derivation of Eq.~(\ref{freecaseineq12}) follows 
along the same lines that led us to~(\ref{DER1}) and (\ref{DER2}).
Specifically  we can write 
\begin{eqnarray}  
&&\ergo_S (\rh_{SE}, \ham_{SE})  \\
\nonumber && \qquad  \geq\ergo_S^{\rm off}  (\rh_{SE}, \ham_{SE})  - \mbox{Tr}\left[
\hat{U}^{(\rm{free})}_S \rh_{SE}\hat{U}^{(\rm{free})^\dag}_S\hat{V}_{SE}\right] \\
\nonumber && \qquad  \geq\ergo_S^{\rm off}  (\rh_{SE}, \ham_{SE})  - \|  \rh_{SE}\|_2 
\|\hat{V}_{SE}\|_2 \;, 
\end{eqnarray} 
where  $\hat{U}^{(\rm{free})}_S$ is the
optimal unitary associated with the free model scenario~(\ref{freecase}).
Similarly we can write 
\begin{eqnarray} 
&& \ergo_S (\rh_{SE}, \ham_{SE}) \\
&&\qquad \leq \ergo_S^{\rm off}  (\rh_{SE}, \ham_{SE}) - \min_{\hat{U}_S \in \mathbf{U}(d_S)}\mbox{Tr}\left[\hat{U}_S\rh_{SE}\hat{U}_S^\dag \hat{V}_{SE}\right]\nonumber \\
&&\qquad \leq \ergo_S^{\rm off}  (\rh_{SE}, \ham_{SE}) + \|  \rh_{SE}\|_2 
\|\hat{V}_{SE}\|_2\;. \nonumber 
  \end{eqnarray}

\section{Generalized Bloch vectors}\label{sec:prelim}
Assuming that  the Hilbert space $\hilb_{S}$ of  $S$ has finite dimension $d_S$
a GPO set
is a collection $\{ \hat\sigma_{S}^{(i)}; i=1,\cdots, d^2-1\}$ of
$(d^2-1)$ self-adjoint operators 
that fulfil the properties
	\begin{eqnarray}
	\mbox{Tr}[\hat\sigma_{S}^{(i)}] = 0 \;,  \label{traccia_pauli}  \qquad 
	\mbox{Tr}[\hat\sigma_{S}^{(i)} \hat\sigma_{S}^{(j)}] = 2 \delta_{ij} \;.
	\end{eqnarray}
Together with the identity operator $\hat{I}_S$ a GPO set form a basis for the operators $\hat \theta_S$ on $\hilb_{S}$ which leads to the following  expansion formula 
	\begin{eqnarray} \label{expansion} 
	\hat\theta_S = \frac{\mbox{Tr}[ \hat\theta]}{d_S}   \; \hat{I}_S +\frac{1}{2} \sum_{i=1}^{d^2-1} q_i(\hat{\theta}_S) \; \hat\sigma_{S}^{(i)} \;, 
	\end{eqnarray}
	where for $i\in\{ 1,\cdots, d_S^2-1\}$ the coefficients 
	\begin{eqnarray}
	q_i(\hat\theta_S) := \mbox{Tr}[\hat\theta_S\hat\sigma_{S}^{(i)}] \; ,
	\label{generalizedBlochCoordinates}
	\end{eqnarray}
	are the complex components of a $(d_S^2-1)$-dimensional vector $\vec{q}(\hat\theta_S)$ whose norm corresponds to 
  Hilbert-Schmidt norm
	of $\hat{\theta}_S$ up to a scaling  factor
	\begin{eqnarray}
	 \left| \vec{q}(\hat\theta_S)\right| = \| \hat\theta_S - \Tr[\theta_S] \hat{I}_S \|_2 / \sqrt{2}\;. 
	\end{eqnarray}
	Equation~(\ref{decomposizione_hamiltoniana_loc}) 
	is a direct application of (\ref{expansion}), while Eqs.~(\ref{decomposizione_statooperatore}) and ~(\ref{decomposizione_hamiltoniana}) follow 
	from a trivial generalization of such identity to the case of a joint operator $\hat\theta_{SE}$ of $S$ and $E$. 
	We also recall that given $\{ | j\rangle_S\}_{j=0,\cdots,d_S-1}$ an orthonormal basis  of $\hilb_S$
	  a special example of GPOs is provided by 
the  matrices
	\begin{eqnarray}
	\hat{\sigma}^x_{S,jj'} &:=&  |j\rangle_S\langle j'|  +  |j'\rangle_S\langle j|  \;, \label{lambdax} \\
	\hat{\sigma}^y_{S,jj'} &:=& i( |j'\rangle_S\langle j| -|j\rangle_S\langle j'|) \;, 
	\nonumber  
	\\
	\hat{\sigma}^z_{S,k} &:=& \sqrt{\tfrac{2}{d_S(d_S-1)}} \Big( (1-d_S+k) |0\rangle_S\langle 0| \nonumber \\
	&& + \sum_{j=1}^{d_S-k-1} |j\rangle_S\langle j|  \Big) \;, 
	\nonumber 
	\end{eqnarray}
	for $0 \leq j < j' \leq d_S-1$ and $0 \leq k \leq d_S-2$.
	 In the case of a qubit (i.e., $d_S=2$) this choice 
	 leads to the 
	  \emph{Bloch vector} representation  \cite{Bloch1946} which induces a one-to-one correspondence between the quantum states $\rh_S$  and 
	  the unitary ball of $\mathbb{R}^3$ via the mapping 
	\begin{equation}
	\vec{r}(\rh_S) = 
	 \left( \mbox{Tr}[\rh_S\hat{\sigma}_{S}^{x}] , \mbox{Tr}[\rh_S\hat{\sigma}_{S}^{y}], \mbox{Tr}[\rh_S \hat{\sigma}_{S}^{z}]  \right) \; ,
	\label{blochvector}
	\end{equation}
	with $\hat{\sigma}_{S}^{x}:= \sigmax_{S,01}$, $\hat{\sigma}_{S}^{y}:= \sigmay_{S,01}$, and $\hat{\sigma}_{S}^{z}:=\sigmaz_{S,0}$ being the standard Pauli operators. 
	We can also associate (up to a phase factor) to any unitary matrix $\hat{U}_S\in {\bf U}(2)$ an orthogonal matrix ${\cal O}_U \in {\bf O}(3)$ such that, for every state~$\rh_S$
	we get 
	\begin{eqnarray}
	\vec{r} \big( \hat{U}_S \rh_S \hat{U}_S^\dagger \big) = {\cal O}_U	\vec{r} \left(\rh_S\right) \; .
	\label{unitarie_diventanorotazioni}
	\end{eqnarray} 
	For dimension $d_S > 2$, we can still define a generalized Bloch vector \cite{Hioe1981} $\vec{r}(\rh_S) \in \mathbb{R}^{d_S^2 - 1}$, with coordinates as in~(\ref{generalizedBlochCoordinates}), and it is still true that to any unitary transformation in the Hilbert space corresponds an orthogonal transformation in the Bloch space verifying~(\ref{unitarie_diventanorotazioni}). In this case 
	however it is no longer true that any vector in the unitary ball $\lvert \vec{r} \rvert \leq 1$ can be associated with a physical state~\cite{Kimura2003}, and similarly not all the orthogonal matrices $O \in {\bf O}(d_S^2-1)$ are associated to unitary transformations $\hat{U}_S \in {\bf U}(d_S)$ via~(\ref{unitarie_diventanorotazioni}). 
	However, it is known that if
	\begin{eqnarray} \lvert \vec{r} \rvert \leq 2/{d_S}\;,
	\end{eqnarray}  there exists surely a legitimate quantum state $\rh_S$ associated to the vector $\vec{r}$~\cite{Kimura2005}.

	We finally observe that if also the dimension $d_E$ of the system $E$ is  finite one can also 
	adopt a GPO decomposition for its elements~\cite{Hioe1981,Kimura2003}. In this case, defining
	\begin{eqnarray} 
	r_i&:=& \mbox{Tr}[ \hat\sigma_{S}^{(i)} \rh_S]\;, \qquad  q_i:= \mbox{Tr}[ \hat\sigma_{E}^{(i)} \rh_E]\;, \\
	t_{ij} &:=& \mbox{Tr}\left[ \rh_{SE} \left( \hat\sigma_{S}^{(i)} \otimes \hat\sigma_{E}^{(j)} \right)  \right]\;, \\
	v_{ij} &:=& \mbox{Tr}\left[ \hat{V}_{SE} \left( \hat\sigma_{S}^{(i)} \otimes \hat\sigma_{E}^{(j)} \right)  \right]/4\;,
	\end{eqnarray} 
	we can write	\begin{eqnarray}
	\rh_{SE} &=& \frac{1 }{d_S d_E}   \hat{I}_{S} \otimes  \hat{I}_{E}
	 + \frac{1}{4}\sum_{i=1}^{d^2_S-1} \sum_{j=1}^{d^2_E-1} t_{ij}\; \hat\sigma_{S}^{(i)} \otimes \hat\sigma_{E}^{(j)}   \nonumber 
	 \\ \nonumber 
	&& + \frac{1}{2}\sum_{i=1}^{d_S} r_i \; \hat\sigma_{S}^{(i)} \otimes  \hat{I}_{E}  +\frac{1}{2} \sum_{j=1}^{d^2_E-1}  q_{j} \;  \hat{I}_{S} \otimes\hat\sigma_{E}^{(j)} \; ,
	\\ 
\label{decomposizione_hamiltoniana_loc_new}
\hat{V}_{SE} &=& 
\sum_{i=1}^{d^2_S-1} \sum_{j=1}^{d^2_E-1} v_{ij} \; \hat\sigma_{S}^{(i)}  \otimes  \hat\sigma_{E}^{(j)}  \;, 
\end{eqnarray} 
which leads to a rewriting of Eq.~(\ref{def_minf}) as
\begin{eqnarray}
\label{def_m}
\mathcal{M}_{ik} = - \left( r_i h_{k}+ \sum_{j=1}^{d_E^2-1}  t_{ij}v_{kj}\right)\;.
\end{eqnarray}

\section{Two-level Hamiltonian with pure input state}\label{fine} 
A lower bound for $\ergo_S (\rh_{ SE}, \ham_{SE})$ can be established by
 rewriting Eq.~(\ref{def_localergotropy}) as
\begin{eqnarray}
\label{localergotropy2}
&&\ergo_S (\rh_{ SE}, \ham_{SE}) 
= \max_{\hat{U}_S \in \mathbf{U}(d_S)} \left\{ \mbox{Tr}[\ham_{S} ( \rh_{S} - \hat{U}_S \rh_{S} \hat{U}^\dagger_S ) ]  \right. \nonumber \\ \label{scomposizione_ergo}
&& \; + \left. \mbox{Tr}[\hat{V}_{SE} ( \rh_{SE} - (\hat{U}_S \otimes \hat{I}_{E}) \rh_{SE} (\hat{U}_S^\dagger  \otimes \hat{I}_{E})) ] \right\}\;,
\end{eqnarray}
 which, thanks to   the spectral decompositions
$\ham_S + \hat{V}_{SE} = \sum_{k=0}^{d_Sd_E} \epsilon_k \ket{\epsilon_k}_{SE}\bra{\epsilon_k}$ 
and 
$\rh_{SE} = \sum_{j=0}^{d_Sd_E} p_j \ket{j}_{SE}\bra{j}$, 
can be casted in the form 
\begin{eqnarray}
\label{localergotropy4}
&&\ergo_S\left(\rh_{SE} , \ham_{SE} \right) = \Tr[\rh_{SE}(\ham_S + \hat{V}_{SE})]  \nonumber \\
&&\qquad +\max_{\hat{U}_S \in \mathbf{U}(d_S)} \sum_{k,j=1}^{d_Sd_E} -p_j \epsilon_k \left\lvert \Tr\left[ \hat{U}_S \hat{N}^{(k,j)}_{S} \right] \right\rvert^2 \;,
\end{eqnarray}
with $\hat{N}^{(k,j)}_{S} := \Tr_E \left[ \ket{\epsilon_k}_{SE}\bra{j} \right]$.
Next we can notice that thanks to the unitarity of $\hat{U}_S$, we have
\begin{eqnarray}
\max_{\hat{U}_S \in \mathbf{U}(d_S)} \left\lvert \Tr\left[ \hat{U}_S \hat{N}^{(k,j)}_{S} \right] \right\rvert^2 = \| \hat{N}^{(k,j)}_{S} \|^2_1 \; ,
\label{minimo_suuntermine}
\end{eqnarray}
with $\| \cdots\|_1$ being the trace norm symbol. 
Up to a shift in the operator $\ham_S + \hat{V}_{SE}$, we can always assume that $-\epsilon_k \geq 0$ for every $k$. Then, using~(\ref{minimo_suuntermine}) into~(\ref{localergotropy4}), we immediately have the bound
\begin{eqnarray}
\ergo_S\left(\rh_{SE} , \ham_{SE} \right)&\geq& \Tr[\rh_{SE}(\ham_S + \hat{V}_{SE})] \nonumber \\
&&- \sum_{k,j=1}^{d_Sd_E} p_j \epsilon_k \left\| \hat{N}^{(k,j)}_{S} \right\|^2_1 \; .
\label{bound_localergo}
\end{eqnarray}
Furthermore, we know that for every $k$ and $j$ there exists an unitary matrix $\hat{U}^{(k,j)}_{S\ast}$ which saturates the inequality~(\ref{minimo_suuntermine}). Therefore, in the special case in which the state $\rh_{SE} = \ket{\Psi}_{SE}\bra{\Psi}$ is a pure state, and in which the Hamiltonian $\ham_S + \hat{V}_{SE}$ has only two levels and a non-degenerate ground state (so that we can assume, without loss of generality, $\epsilon_1 = -E$ and $\epsilon_2 = 0$), the bound~(\ref{bound_localergo}) becomes an exact equality, and we have
\begin{eqnarray}
&&\ergo_S\left(\ket{\Psi}_{SE} , \ham^{\rm{(two-levels)}}_{SE} \right)  \nonumber \\ &&\qquad =\Tr[\rh_{SE}(\ham_S + \hat{V}_{SE})] - E \left\| \hat{N}^{(1,1)}_{S} \right\|^2_1 \; .
\label{bound_localergo_casospeciale}
\end{eqnarray}


\end{document}